\RequirePackage{fix-cm}
\documentclass[5p,twocolumn,sort&compress]{elsarticle}
\usepackage{graphicx}
\usepackage{siunitx}
\usepackage{tikz}
\usetikzlibrary{arrows,arrows.meta,positioning,calc,math}
\tikzset{
 arrow/.style={-latex, shorten >=1ex, shorten <=1ex}
 }
\usepackage{pgfplots}
\pgfplotsset{compat=1.16}
\usepackage{geometry}

\usepackage{xcolor}
\definecolor{azure}{rgb}{0.0, 0.5, 1.0}
\definecolor{awesome}{rgb}{1.0, 0.13, 0.32}
\definecolor{asparagus}{rgb}{0.53, 0.66, 0.42}
\definecolor{cadetgrey}{rgb}{0.57, 0.64, 0.69}

\usepackage{xfrac}

\usepackage{booktabs}

\usepackage[hidelinks]{hyperref}

\usepackage{mathptmx}      
\usepackage{amssymb}
\usepackage{amsmath}
\usepackage{bm}
\usepackage{nicefrac}
\usepackage[section]{placeins}

\usepackage{subfig}

\usepackage{algorithmicx}
\usepackage{algpseudocode}
\usepackage{algorithm}
\usepackage{moresize}

\usepackage{romannum}
\AtBeginDocument{\pagenumbering{arabic}}

\newcommand{\xb}{\mathbf{x}}
\newcommand{\yb}{\mathbf{y}}
\newcommand{\fb}{\mathbf{f}}
\newcommand{\ub}{\mathbf{u}}

\newcommand{\bb}{\mathbf{b}}

\newcommand{\Bspan}[1]{\mathrm{span}\big\langle{#1}\big\rangle}

\newcommand{\traction}{\bar{\mathbf{t}}}

\newcommand{\mySubScriptSize}{\ssmall}
\newcommand{\hPum}{h_\text{\mySubScriptSize{PUM}}}
\newcommand{\hPd}{h_\text{\mySubScriptSize{PD}}}

\journal{arxiv.org}

\begin{document}

\begin{frontmatter}
\title{A Multiscale Fracture Model using Peridynamic Enrichment of Finite Elements within an Adaptive Partition of Unity: Experimental Validation}

\author[ins,scai]{Matthias Birner}
\ead{mbirner@posteo.net}

\author[cct,lsu+phys]{Patrick Diehl\corref{mycorrespondingauthor}}
\cortext[mycorrespondingauthor]{Corresponding author}
\ead{pdiehl@cct.lsu.edu}
\ead[url]{http://orcid.org/0000-0003-3922-8419}

\author[lsu]{Robert Lipton}
\ead{lipton@lsu.edu}

\author[ins,scai]{Marc Alexander Schweitzer}
\ead{schweitzer@ins.uni-bonn.de}

\address[ins]{Institute for Numerical Simulation, University of Bonn, Bonn, Germany}

\address[scai]{Fraunhofer SCAI, Sankt Augustin, Germany}

\address[cct]{Center for Computation \& Technology, Louisiana State University, Baton Rouge, LA}

\address[lsu]{Department of Mathematics, Louisiana State University, Baton Rouge, LA}

\address[lsu+phys]{Department of Physics and Astronomy, Louisiana State University, Baton Rouge, LA}


\begin{abstract}
Partition of unity methods (PUM) are of domain decomposition type and provide the opportunity for multiscale and multiphysics numerical modeling. Within the PUM global-local enrichment scheme \cite{duarte2007global, birner2017global} different physical models can exist to capture multiscale behavior. For instance, we consider classical linear elasticity globally and local zones where fractures occur. The elastic fields of the undamaged media provide appropriate boundary data for local PD simulations on a subdomain containing the crack tip to grow the crack path. Once the updated crack path is found the elastic field in the body and surrounding the crack is updated using PUM basis with appropriate enrichment near the crack. The subdomain for the PD simulation is chosen to include the current crack tip as well as nearby features that will influence crack growth. 
This paper is part \Romannum{2} of this series and validates the combined PD/PUM simulator against the experimental results presented in~\cite{ingraffea1990probabilistic}.
The presented results show that we can attain good agreement between experimental and simulation data with a local PD subdomain that is moving with the crack tip and adaptively chosen size.
\end{abstract}


\end{frontmatter}



\author{Matthias Birner \and
        Patrick Diehl \and
        Robert Lipton \and
        Marc Alexander Schweitzer
}




\section{Introduction}

This paper provides the experimental validation of \cite{BIRNER2023103360} which is a new method to overcome the computational expense of Peridynamic (PD) discretization, by using the partition of unity (PU) approach \cite{schweitzer2003phd}. 
The overall process is based on the construction of a combined PD/PU simulator which automatically determines the regions where a local PD model and a classical linear elastic model should be employed so that the resulting local PD approximation can be utilized in building a multiscale enrichment function for the global partition of unity method (PUM). This provides a numerical approximation that captures the true material response including fracture growth with high accuracy efficiently.
In this paper, we compare our method to experimental results to validate the approach.
Namely, we leverage the compatibility of the PD and classical material models and the methods required for combining them. To accomplish this, we seek to outline the transfer of information between them. The use of elastic field boundary conditions to drive crack growth is aided using the PD model introduced in \cite{lipton2014dynamic, lipton2016cohesive}. The energy for this model interpolates between linear elastic energy for small strains and surface energy associated with displacement jumps. It sidesteps the surface effects associated with nonscaled standard peridynamic formulations. The subdomain needed for the PD simulation is chosen to include the current crack tip together with nearby features that will influence crack growth.  Once the current crack geometry is established via the local PD approximation we construct respective enrichment basis functions and use the PUM to efficiently determine the elastic displacement outside the crack and in the complete body. This framework accomplishes a so-called global-local enrichment strategy, see \cite{duarte2007global, birner2017global}. We demonstrate an adaptive strategy for choosing the PD subdomain that allows for good agreement between experiment and simulation. It is found that larger fixed PD subdomains result in less accurate crack paths than using the smaller adaptive PD subdomains proposed in our approach.

Several approaches to combine the PD model with discretizations of linear elasticity already exist, see~\cite{fem:pd:coupling:review} for a review.
Most similar to our approach, due to also leveraging a partition of unity method, is the coupling method with the XFEM presented in~\cite{xfem:pd:coupling}.
In all the above methods, however, an interface between the involved models and discretizations is introduced, which can cause problems~\cite{fem:pd:coupling:waves}.
The proposed global-local enrichments based coupling sidesteps this issue by separating the models on independent discretizations, i.e. our approach is essentially a hierarchical coupling technique.

The paper is structured as follows: Section~\ref{sec:preliminaries} recaps briefly the preliminaries of the classical continuum mechanics, partition of unity method, and peridynamic theory. For a more detailed description, we refer to the Part \Romannum{1} of the paper series~\cite{BIRNER2023103360}. In Section~\ref{sec:numerical:results} the combined PUM/PD simulator is validated against three experiments. Finally, Section~\ref{sec:conclusion} concludes the paper with some remarks on missing features to fully automate the simulator.

\section{Preliminaries}
\label{sec:preliminaries}
We provide a brief recap of the two methods in the combined simulator, the PUM and PD, and state the equations we solve with each, the equations of linear elasticity, and the peridynamic material model.
For a more detailed description, we refer to Part I of the series~\cite{BIRNER2023103360}.


\subsection{Partition of Unity method}
In a partition of unity method~\cite{melenk1996partition, babuvska1997partition,duarte2000generalized,moes1999finite}, local approximation spaces \(\mathrm{V}_i\) are attached to PU functions \(\varphi_i \geq 0\) that cover the computational domain \(\Omega\).
The resulting finite-dimensional ansatz space \(\mathrm{V}^{PU} = \sum_i \varphi_i \mathrm{V}_i\) is then used in the Galerkin method to discretize a PDE.
Usually, the local spaces consist of polynomial functions \(\bm{\psi}_i^s\) plus problem specific basis functions \(\bm{\eta}_i^t\), so called enrichments
\begin{equation}
    \mathrm{V}_i = \mathcal{P}_i + \mathcal{E}_i  = \Bspan{\bm{\psi}_i^s, \bm{\eta}_i^t}.
\end{equation}
Enrichments can be analytically given functions that are known to capture local characteristics of the solution as well as numerical solutions to other simulations.
Note that these two spaces $\mathcal{P}_i$ and $\mathcal{E}_i$ can be selected independently at each PU function and the overall ansatz space is
\begin{equation}\label{eq:def_pu_space}
    \mathrm{V}^{PU} := \sum_i \varphi_i \mathrm{V}_i = \sum_i \varphi_i \mathcal{P}_i + \varphi_i \mathcal{E}_i.
\end{equation}
%

In this publication, we employ the specific PUM presented in~\cite{schweitzer2003phd} and run all experiments with the PUMA software framework~\cite{schweitzer2017rapid}.
Compared to other popular finite element based partition of unity methods, the PUM is a meshfree method and has the advantage that the construction of the underlying PU allows to guarantee a stable basis independent of the applied enrichments.

In this paper, we solve the equation of linear elasticity with the PUM, which is given by 
\begin{equation}\begin{aligned}\label{eq:lin:elasticity}
    \nabla \cdot \bm{\sigma} &= \bb \qquad &\mathrm{in} \; &\Omega \\
    \ub(\xb) &= \bar{\ub}(t, \xb) \quad &\mathrm{on}\; & \Gamma^D \\
    \bm{\sigma}(\xb) \cdot \mathbf{n}(\xb) &= \traction(\xb) \quad &\mathrm{on}\; &\Gamma^N
\end{aligned}\end{equation}
on a domain \(\Omega\) with boundary \( \partial\Omega = \Gamma_N \mathbin{\dot{\cup}} \Gamma_D \).
Here, \(\mathbf{n}\) is the normal on the boundary, \(\bb\) represent volume forces acting on the body and \(\bm{\sigma}\) is the Cauchy stress tensor given by
\begin{equation}\label{eq:Hookes_law}
    \bm{\sigma} = \mathbf{C} : \bm{\varepsilon} = 2\mu\bm{\varepsilon}(\ub) + \lambda \mathrm{tr}\left(\bm{\varepsilon}(\ub)\right) \bm{\mathbb{I}} ,
\end{equation}
for linear and isotropic media.
The PUM discretizes the weak formulation of~\eqref{eq:lin:elasticity}.

\subsection{Peridynamics}
Peridynamics (PD)~\cite{silling2000reformulation,silling2005meshfree}, a non-local generalization of classical continuum mechanics (CCM) which allows for discontinuous displacement fields. PD provides an attractive framework for simulating growing cracks and fractures and was employed in many validations against experimental data~\cite{diehl2019review}.

In PD the equation of motion is given by
\begin{equation}\label{pd:cont:equation}
    \rho(\xb) \ddot{\ub}(\xb, t) = \int_{B_{\delta}(\xb)} \fb\big(\yb - \xb,\, \ub(\yb, t) - \ub(\xb, t)\big) \,\mathrm{d}\yb + \bb(\xb, t),
\end{equation}
where \(\rho\) is the material density, \(\ub\) the displacement, \(\bb\) the external force density and the pair-wise force density \(\fb\) encodes the PD material model.
In this paper we use the bond-based softening model presented in~\cite{pd:material:1, pd:material:2}, where the pair-wise force density
\begin{equation}\label{eq:pd:def:force:density}
    \fb(\Delta\xb, \Delta \ub) :=  \frac{\partial_S \psi\left( \Delta\xb, \Delta \ub \right)}{|\Delta\xb|} e_{\Delta\xb}
\end{equation}
is given by the derivative of the pair-wise force potential \(\psi\) with respect to the bond stretch \(S\) and
\begin{equation}
    e_{\Delta\xb} := \frac{\yb - \xb}{|\yb - \xb|}
\end{equation}
is a unit vector.
The bond stretch \(S\) between two points \(\xb, \yb\) is determined by their difference \(\Delta\xb = \yb - \xb\) and the difference between their respective displacements \(\Delta \ub = \ub(\yb, t) - \ub(\xb, t)\) by
\begin{equation}
    S(\Delta\xb, \Delta \ub) := \frac{\Delta \ub}{|\Delta\xb|} \cdot e_{\Delta\xb} = \frac{\ub(\yb, t) - \ub(\xb, t)}{|\yb - \xb|}  \cdot \frac{\yb - x}{|\yb - \xb|}.
\end{equation}
The bond stretch derivative of the pair-wise force potential
\begin{equation}
    \partial_S \psi(\Delta\xb, \Delta \ub) = J^\delta\left(|\Delta\xb|\right) \frac{\partial_S g\left(|\Delta\xb| \, S^2(\Delta\xb, \Delta \ub)\right)}{\delta \, \mu\left(B_{\delta}(0)\right)}
\end{equation}
is defined by the piecewise constant influence function
\begin{equation}
    J^\delta(r) := \left\{
        \begin{array}{lll}
            1  & \mbox{if }& 0 \leq r < \delta \\
            0  & \mbox{else }&
        \end{array}
    \right.
\end{equation}
and the double well potential
\begin{equation}\label{eq:pd:def:double:well}
    g(r) := C (1 - \exp[-\beta r]),
\end{equation}
with \(C\) and \(\beta\) material parameters.
For a given critical energy release rate \(G_c\), Young's modulus \(E\) and Poisson's ration \(\nu\), the consistent PD material parameters are given by
\begin{equation}
    C := \pi \frac{G_c}{4} \quad \text{and} \quad \beta := \frac{4 E \nu}{C (1 - \nu)(1 - 2\nu)}.
\end{equation}
Note, however, that in bond-based peridynamics Poisson's ratio is limited to \(\nicefrac{1}{3}\) for plain-strain, hence we fix $\nu=\nicefrac{1}{3}$ so that our material model is defined by the energy release rate and Young's modulus only.

\section{Combined PUM/PD simulator}
\label{sec:simulator}

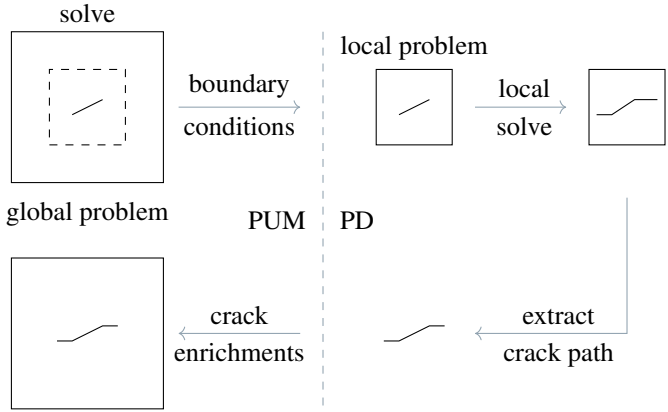
\begin{figure}[tb]
    \centering
    \begin{tikzpicture}

        \draw (0,3) -- (2,3) -- (2,5) -- (0,5) -- cycle;
        \draw[dashed] (0.5,3.5) -- (1.5,3.5) -- (1.5,4.5) -- (0.5,4.5) -- cycle;
        \draw (0.8,3.9) -- (1.2,4.1);
        \node[above] at (1,2.3) {global problem};
        \node[above] at (1,5) {solve};
        \node[above] at (3,4) {boundary};
        \draw[thin,cadetgrey,->] (2.2,4) -- (3.8,4) ;
        \node[below] at (3,4) {conditions};
        \node[right] at (4.2,2.5) {PD};
        \node[above] at (5.3,4.5) {local problem};
        \draw (4.8,3.5) -- (5.8,3.5) -- (5.8,4.5) -- (4.8,4.5) -- cycle;
        \draw (5.1,3.9) -- (5.5,4.1);

        \node[above] at (6.75,4) {local};
        \draw[thin,cadetgrey,->] (6.1,4) -- (7.3,4) ;
        \node[below] at (6.75,4) {solve};
        \draw (7.6,3.5) -- (8.6,3.5) -- (8.6,4.5) -- (7.6,4.5) -- cycle;
        \draw (7.7,3.9) -- (7.9,3.9) -- (8.2,4.1) -- (8.5,4.1);
        \node[above] at (7.2,1) {extract};
        \draw[thin,cadetgrey,->] (8.1,2.8) -- (8.1,1) -- (6.1,1) ;
        \node[below] at (7.2,1) {crack path};

        \draw (4.9,0.9) -- (5.1,0.9) -- (5.5,1.1) -- (5.7,1.1);
        \node[left] at (4,2.5) {PUM};
        \node[above] at (3,1) {crack};
        \draw[thin,cadetgrey,<-] (2.2,1) -- (3.8,1) ;
        \node[below] at (3,1) {enrichments};
        \draw (0,0) -- (2,0) -- (2,2) -- (0,2) -- cycle;
        \draw (0.6,0.9) -- (0.8,0.9) -- (1.2,1.1) -- (1.4,1.1);
        \draw[thin, dashed, cadetgrey] (4.1,5) -- (4.1,0);
    \end{tikzpicture}
    \caption{The combined PUM/PD simulator. Adapted from~\cite{BIRNER2023103360}.}
    \label{fig:coupling:glcycle}
\end{figure}

We combine the PUM with PD using the global-local enrichment method~\cite{duarte2007global, birner2017global} as sketched in Figure~\ref{fig:coupling:glcycle} and introduced in more detail in~\cite{diss:Birner}.
In this approach, we first solve the global linear elasticity problem with the PUM to provide the computed displacement as boundary data for the local PD problem.
We extract the updated crack path from the local PD solution and model it with standard crack enrichments~\cite{birner2017global} in the global PUM simulation, which we solve again.

The algorithm sketched so far ignores time.
In a dynamic simulation, we would run both methods concurrently and exchange information at some frequency, where having smaller time steps on the local PD problem is probably advised.
We did not yet investigate the combined method in time, however, the standard global-local method has already been applied successfully in a dynamic simulation~\cite{gl:in:time}.
In this paper, we simulate all examples quasi-statically and update the boundary data and the crack path at certain load steps.
A quasi-static PD solution of the problems at hand would have been computationally too expensive and we instead approximate it by slowly scaling the applied displacement in time on the local PD problem.
Moreover, we currently extract the PD crack paths by hand and plan to investigate automatic crack path extraction in future publications.

\section{Numerical results}
\label{sec:numerical:results}

\begin{figure*}[tb]
    \centering
\begin{tikzpicture}

\draw (0,0) -- (9,0) -- (9,4) -- (0,4) -- cycle;

\draw[thick,->] (4.5,4.5) -- (4.5,4);
\node[above] at (4.5,4.5) {$u$};
\draw[->,thick,cadetgrey] (-2.5+1,0) -- (-2+1,0);
\draw[->,thick,cadetgrey] (-2.5+1,0) -- (-2.5+1,0.5);
\node[above] at (-2.5+1,0.5) {\small $y$};
\node[right] at (-2+1,0.) {\small $x$};
\draw[thin,cadetgrey] (9,0) -- (9.5,0);
\draw[thin,cadetgrey] (9,4) -- (9.5,4);
\draw[thin,cadetgrey,<->] (9.45,0) -- (9.45,4);
\node[right] at (9.5,2) {8};
\draw (9/4,4-1.5/2) circle (0.125);
\draw (9/4,4-3.5/2) circle (0.125);
\draw (9/4,4-5.5/2) circle (0.125);
\draw[thin,cadetgrey] (9/4+0.25,4-5.5/2) -- (9/4+0.5,4-5.5/2);
\draw[thin,cadetgrey] (9/4+0.25,4-3.5/2) -- (9/4+0.5,4-3.5/2);
\draw[thin,cadetgrey] (9/4+0.25,4-1.5/2) -- (9/4+0.5,4-1.5/2);
\draw[thin,cadetgrey,<->] (9/4+0.375,4-5.5/2) -- (9/4+0.375,4-3.5/2);
\draw[thin,cadetgrey,<->] (9/4+0.375,4-3.5/2) -- (9/4+0.375,4-1.5/2);
\draw[thin,cadetgrey,<->] (9/4+0.375,4-1.5/2) -- (9/4+0.375,4);
\node[right] at (9/4+0.375,4-1.5/4+0.125) {1.25};
\node[right] at (9/4+0.375,4-1.5/2-0.5) {2};
\node[right] at (9/4+0.375,4-3.5/2-0.5) {2};
\draw[thick] (9/5,0) -- (9/5,0.5);
\node[right] at (9/5+0.5,0.25) {$a$};
\draw[cadetgrey,thin] (9/5,0.5) -- (9/5+0.5,0.5);
\draw[thin,cadetgrey,<->] (9/5+0.45,0) -- (9/5+0.45,0.5);
\foreach \j in {5,...,5}
        \foreach \i in {-0.5,-0.5}
        {
        \draw (0.25+2.5-0.5*\j,0.24+\i*0.5) -- (0.1+2.5-0.5*\j,-0.15+0.25+\i*0.5);
        \draw (0.25+2.5-0.5*\j,0.24+\i*0.5) -- (0.4+2.5-0.5*\j,-0.15+0.25+\i*0.5);
        \draw (0.1+2.5-0.5*\j,-0.15+0.25+\i*0.5) -- (0.4+2.5-0.5*\j,-0.15+0.25+\i*0.5);
        \draw (0.03+2.5-0.5*\j,-0.25+\i*0.5+0.35) -- (0.45+2.5-0.5*\j,-0.25+\i*0.5+0.35);
        \draw (0.4+2.5-0.5*\j,-0.25+0.35+\i*0.5) -- (0.3+2.5-0.5*\j,-0.3+0.35+\i*0.5);
        \draw (0.3+2.5-0.5*\j,-0.25+0.35+\i*0.5) -- (0.2+2.5-0.5*\j,-0.3+0.35+\i*0.5);
        \draw (0.2+2.5-0.5*\j,-0.25+0.35+\i*0.5) -- (0.1+2.5-0.5*\j,-0.3+0.35+\i*0.5);
        }
\foreach \j in {-12,...,-12}
        \foreach \i in {-0.5,-0.5}
        {
        \draw (0.25+2.5-0.5*\j,0.24+\i*0.5) -- (0.1+2.5-0.5*\j,-0.15+0.25+\i*0.5);
        \draw (0.25+2.5-0.5*\j,0.24+\i*0.5) -- (0.4+2.5-0.5*\j,-0.15+0.25+\i*0.5);
        \draw (0.1+2.5-0.5*\j,-0.15+0.25+\i*0.5) -- (0.4+2.5-0.5*\j,-0.15+0.25+\i*0.5);
        \draw (0.03+2.5-0.5*\j,-0.25+\i*0.5+0.35) -- (0.45+2.5-0.5*\j,-0.25+\i*0.5+0.35);
        \draw (0.4+2.5-0.5*\j,-0.25+0.35+\i*0.5) -- (0.3+2.5-0.5*\j,-0.3+0.35+\i*0.5);
        \draw (0.3+2.5-0.5*\j,-0.25+0.35+\i*0.5) -- (0.2+2.5-0.5*\j,-0.3+0.35+\i*0.5);
        \draw (0.2+2.5-0.5*\j,-0.25+0.35+\i*0.5) -- (0.1+2.5-0.5*\j,-0.3+0.35+\i*0.5);
        }
\draw[thin,cadetgrey] (0,-0.25) -- (0,-0.75);
\draw[thin,cadetgrey] (9,-0.25) -- (9,-0.75);
\draw[->,thin,cadetgrey] (0,-0.35) -- (0.45,-0.35);
\node[right] at (0.45,-0.35) {1};
\draw[->,thin,cadetgrey] (9,-0.35) -- (9-0.45,-0.35);
\node[left] at (9-0.45,-0.35) {1};
\draw[<->,thin,cadetgrey] (4.5,-0.25) -- (9/5,-0.25); 
\node[below] at (3,-0.25) {$b$};
\draw[<->,thin,cadetgrey] (0,-0.7) -- (4.5,-0.7); 
\draw[<->,thin,cadetgrey] (4.5,-0.7) -- (9,-0.7); 
\node[below] at (2.25,-0.7) {10};
\node[below] at (4.5+2.25,-0.7) {10};
%
%
%
\end{tikzpicture}
    \caption{Parametrized geometry of the 3-point bending experiment. With Parameters $a$ for the length and $b$ for the position in $x$-direction of the initial crack. The values for the three cases are shown in Table~\ref{tab:geometry:3p:bending}. Adapted from~\cite{ingraffea1990probabilistic}.}
    \label{fig:geometry:3p:bending}
\end{figure*}
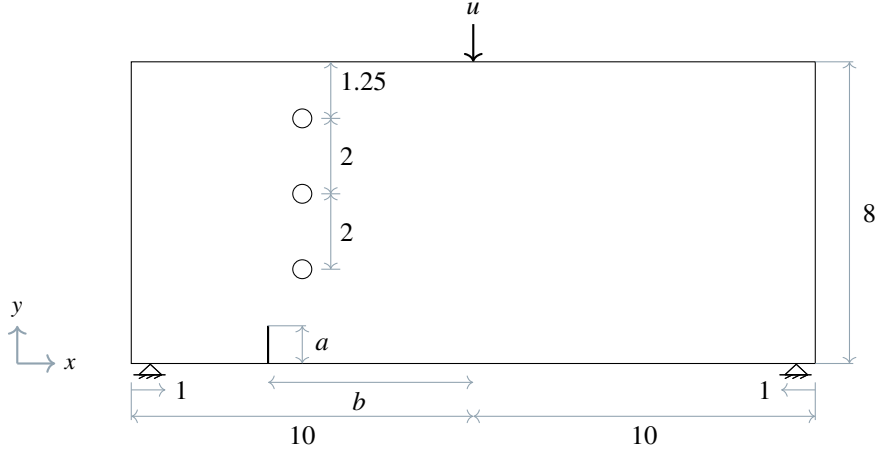

\begin{figure}[tbp]
    \centering
    \includegraphics[width=\linewidth]{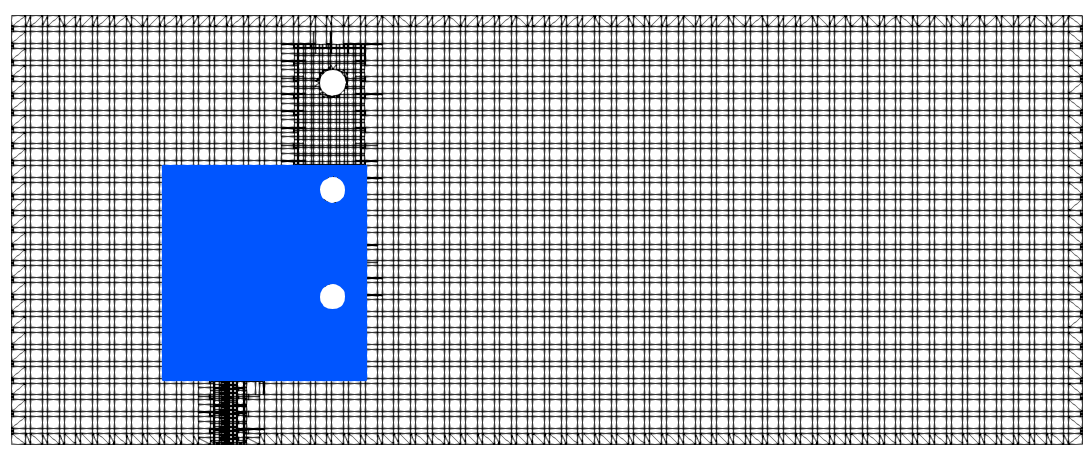}
    \caption{Example for the discretization of the two models: The black mesh is used for the partition of unity method simulation of the global problem for Case \Romannum{2}. The blue dots are the discrete peridynamics nodes within the PD region. For simplicity, we showed one PD region at a later stage of the simulation from Figure~\ref{fig:3bending:crack:path:case2:boxes}.}
    \label{fig:combined:mesh}
\end{figure}

For the comparison against experimental data, we use the three-point bending experiment by Ingraffea et al.~\cite{ingraffea1990probabilistic}. This experiment was used to validate peridynamic simulations~\cite{NI2019126,sheikhbahaei2023efficient}, phase-field simulations~\cite{msekh2015abaqus,wu2018robust,mesgarnejad2015validation,gerasimov2020stochastic}, and XFEM/GFEM simulations~\cite{mukhtar2020validation}. Figure~\ref{fig:geometry:3p:bending} sketches the parametrized geometry with the parameters $a$ the initial length of the crack and $b$ the position in $x$-direction of the crack. Three experiments with the parameters given in Table~\ref{tab:geometry:3p:bending} were conducted. We refer to these as Case \Romannum{1}, Case \Romannum{2}, and Case \Romannum{3} in this paper. Note that the geometry is parameterized in inches, as in the paper~\cite{ingraffea1990probabilistic}. Table~\ref{tab:three:holes:parameters} summarizes the common discrete simulation parameters in time and space employed in this paper. Figure~\ref{fig:combined:mesh} shows in black the mesh for the partition unity method for Case \Romannum{2}. The blue region shows the discrete PD nodes within the PD region. For simplicity, we showcased one of the boxes from Figure~\ref{fig:3bending:crack:path:case2:boxes} at a later simulation stage. \\

The experimental crack path in Figure~\ref{fig:3bending:crack:path:case1}, Figure~\ref{fig:3bending:crack:path:case2}, and Figure~\ref{fig:3bending:crack:path:case3} were extracted from the corresponding Figures 4.8, 5.5, and 5.7 from~\cite{ingraffea1990probabilistic} using the tool \textit{WebPlotDigitizer}~\cite{Rohatgi2022}. We want to mention that some uncertainties are introduced when extracting the crack path coordinates. We compare our results to experimental data as well as to the results presented in~\cite{NI2019126}. There the authors proposed an approach to couple the finite element method with peridynamics considering only quasi-static settings whereas we employ dynamic PD. Moreover, our approach is essentially a hierarchical coupling technique whereas the method in ~\cite{NI2019126} resembles a concurrent coupling approach. Thus, we do not anticipate to observe perfect agreement of our results and those of ~\cite{NI2019126}.

\begin{table}[tb]
    \centering
    \caption{Parameters $a$ for the length and $b$ the position in $x$-direction of the initial crack. The parametrized geometry is shown in Figure~\ref{fig:geometry:3p:bending}. Adapted from~\cite{ingraffea1990probabilistic}.}
    \label{tab:geometry:3p:bending}
   \begin{tabular}{llll}\toprule
        Case & a(in) & b(in) & Number of holes  \\\midrule
        \Romannum{1} & 1 & 6 & 0 \\
        \Romannum{2} & 1 & 6 & 3 \\
        \Romannum{3} & 1.5 & 5 & 3 \\\bottomrule
    \end{tabular}
 \end{table}

\begin{table*}[tb]
    \centering
    \caption{Simulation parameters for the discretization in time and space for Case \Romannum{1}, Case \Romannum{2}, and Case \Romannum{3}.}
    \label{tab:three:holes:parameters}
    \begin{tabular}{l|l}
    \toprule
     Force $F=$\num{9e5}\si{\newton}  & Time steps $t_n$=\num{50000} \\
     Node spacing $\hPd=$\num{0.00049609375}\si{\meter} & Time step size $t_s=$\num{2e-7}\si{\second} \\
     and $\hPum=$\num{0.00396875}\si{\meter} & \\
     Horizon $\delta=8\hPd=$\num{0.00396875}\si{\meter} & Final time $T$=\num{0.001}\si{\second}  \\\bottomrule
    \end{tabular}
\end{table*}

\subsection{Case 1:}

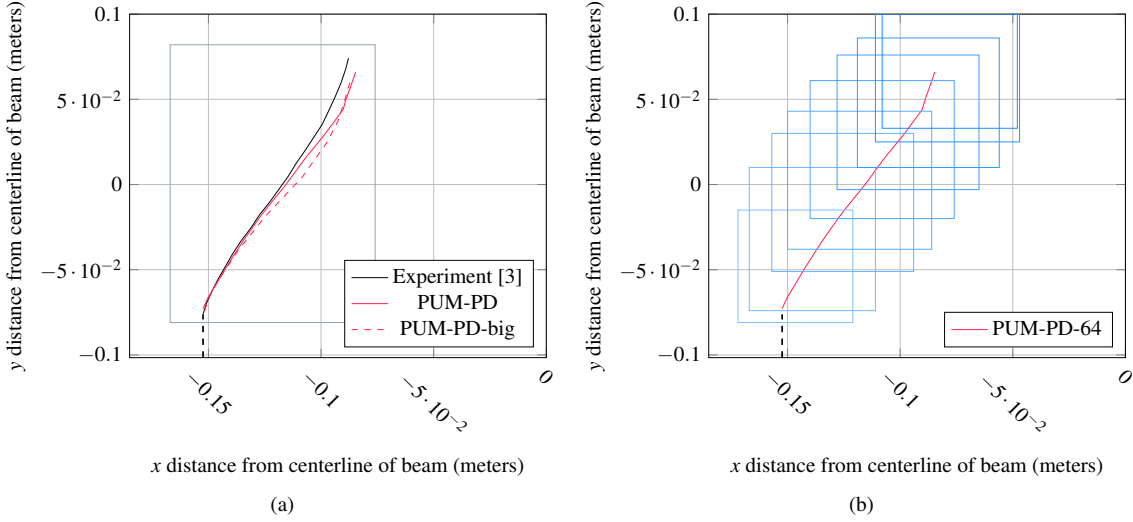
\begin{figure*}[tbp]
    \centering
    \subfloat{
    \begin{tikzpicture}
        \node at (0,0) {Case \Romannum{1}: Number of holes is 0, $a$ = 1, and $b$ = 6 };
    \end{tikzpicture}
    }
    \setcounter{subfigure}{0}
    \\
    \subfloat[\label{fig:3bending:crack:path:case1:path}]{
    \begin{tikzpicture}[scale=0.8]
    \begin{axis}[legend pos=south east, xlabel=$x$ distance from centerline of beam (meters), ylabel=$y$ distance from centerline of beam (meters), grid, ymax=0.1,ymin=-4/39.37,xmax=0, xmin=-0.185,xticklabel style={rotate=-45}]
    \addplot[black,mark=none] table [x expr=-(0.254-\thisrowno{0}), y expr=-(0.1016-\thisrowno{1}), col sep=comma] {3point-case1-reference.csv};
    \addplot[awesome,mark=none] table [x expr=-(0.254-\thisrowno{0}), y expr=-(0.1016-\thisrowno{1}), col sep=comma] {3point-case1-crack-64-birner.csv};
    \addplot[awesome, dashed,mark=none] table [x expr=-(0.254-\thisrowno{0}), y expr=-(0.1016-\thisrowno{1}), col sep=comma] {3point-case1-crack-64-birner-big.csv};
    \draw [thick, draw=black,dashed]
    (axis cs: -6/39.37 ,-4/39.37) -- (axis cs: -6/39.37,-3/39.37);
    \draw[cadetgrey] (-0.167 , -0.081) rectangle (-0.076 , 0.082);
    \legend{Experiment~\cite{ingraffea1990probabilistic}, PUM-PD,PUM-PD-big};
    \end{axis}
    \end{tikzpicture}
    }
    \subfloat[\label{fig:3bending:crack:path:case1:box}]{
    \begin{tikzpicture}[scale=0.8]
    \begin{axis}[legend pos=south east, xlabel=$x$ distance from centerline of beam (meters), ylabel=$y$ distance from centerline of beam (meters), grid, ymax=0.1,ymin=-4/39.37,xmax=0, xmin=-0.185,xticklabel style={rotate=-45}]
    \addplot[awesome,mark=none] table [x expr=-(0.254-\thisrowno{0}), y expr=-(0.1016-\thisrowno{1}), col sep=comma] {3point-case1-crack-64-birner.csv};
    \draw [thick, draw=black,dashed]
    (axis cs: -6/39.37 ,-4/39.37) -- (axis cs: -6/39.37,-3/39.37);
    \draw[azure!50] (-0.172 , -0.081) rectangle (-0.121 , -0.015);
    \draw[azure!55] (-0.167 , -0.074) rectangle (-0.111 , 0.010);
    \draw[azure!60] (-0.157 , -0.051) rectangle (-0.094 , 0.030);
    \draw[azure!65] (-0.150 , -0.038) rectangle (-0.086 , 0.043);
    \draw[azure!70] (-0.140 , -0.020) rectangle (-0.076 , 0.061);
    \draw[azure!75] (-0.128 , -0.003) rectangle (-0.065 , 0.076);
    \draw[azure!80] (-0.119 , 0.010) rectangle (-0.056 , 0.086);
    \draw[azure!85] (-0.111 , 0.025) rectangle (-0.047 , 0.100);
    \draw[azure!90] (-0.108 , 0.033) rectangle (-0.048 , 0.108);
    \legend{PUM-PD-64};
    \end{axis}
    \end{tikzpicture}
    }
    \caption{ \protect\subref{fig:3bending:crack:path:case1:path} Experimentally obtained crack path for the Case \Romannum{1} extracted from Figure 4.8 in~\cite{ingraffea1990probabilistic} in black. The \textcolor{awesome}{red} crack path shows the crack path obtained by our approach. The \textcolor{cadetgrey}{gray} box is the PD domain used in~\cite{NI2019126}. \protect\subref{fig:3bending:crack:path:case1:box} Simulated crack path with our approach in \textcolor{awesome}{red} and the moving PD region with the crack tip position. The \textcolor{azure!50}{light blue} region at the crack tip is the initial square region. After that, the box color moves from \textcolor{azure!50}{light blue} at the beginning to \textcolor{azure!90}{dark blue} for the region used in the last simulation step.}
    \label{fig:3bending:crack:path:case1}
\end{figure*}

For this case, we use first a similar rectangle ($(2,0)$,$(7,7)$) for the PD region as used for the coupling of PD and finite elements in~\cite{NI2019126}, see the gray rectangular in Figure~\ref{fig:3bending:crack:path:case1:path}. The black line shows the crack path obtained by the experiment in~\cite{ingraffea1990probabilistic} in Figure 4.8 on page 49 extracted using \textit{WebPlotDigitizer}. The initial crack is shown as a dashed line.\\

The dashed red line shows the extracted crack path using the same PD region as in~\cite{NI2019126} and a similar nodal spacing. At the beginning, the crack is close to the simulated one with the moving box and the experimental crack path. Later the simulated crack path using the constant bigger box diverges from the other crack path.\\

To reduce the simulation time further, we investigate if a smaller PD region is feasible. Figure~\ref{fig:3bending:crack:path:case1:box} shows the smaller initial squared PD region close to the initial crack in light blue. As the crack grew the initial box moved with the simulated crack tip and the size was adjusted. We plotted a series of the PD region moving with the crack tip in our global domain from light blue to dark blue. The red line in Figure~\ref{fig:3bending:crack:path:case1:path} is the extracted crack path using the moving PD region. At the beginning, the simulated crack path is in good agreement with the experiment. However, after the crack grew around 50\% the simulated crack path started to diverge from the experiment slightly. Over time the difference is advancing while the shape of the crack path is still similar. In addition, the crack tip at the end of the simulation is slightly lower than in the experiment. Similar behavior was observed in~\cite{NI2019126} using the constant larger PD region. Notably, we use a nodal space very close in our simulations to that used in~\cite{NI2019126}. However, our PD region is smaller. Since not all geometric details and parameters for Case \Romannum{1} were given in~\cite{NI2019126}, we could not extract the respective crack paths for direct comparison.

The small differences we observe in Figure~\ref{fig:3bending:crack:path:case1:path} in the attained crack paths most likely can be attributed to the fact that the boundary data exchange is not syncronized in both simulations and the fact that our crack path identification currently is not perfectly robust and fully automatic so that the identified crack paths contain some uncertainty. Here, a more detailed analysis in the future is necessary.

\subsection{Case II:}
\begin{figure*}
    \centering
    \includegraphics[width=0.85\linewidth]{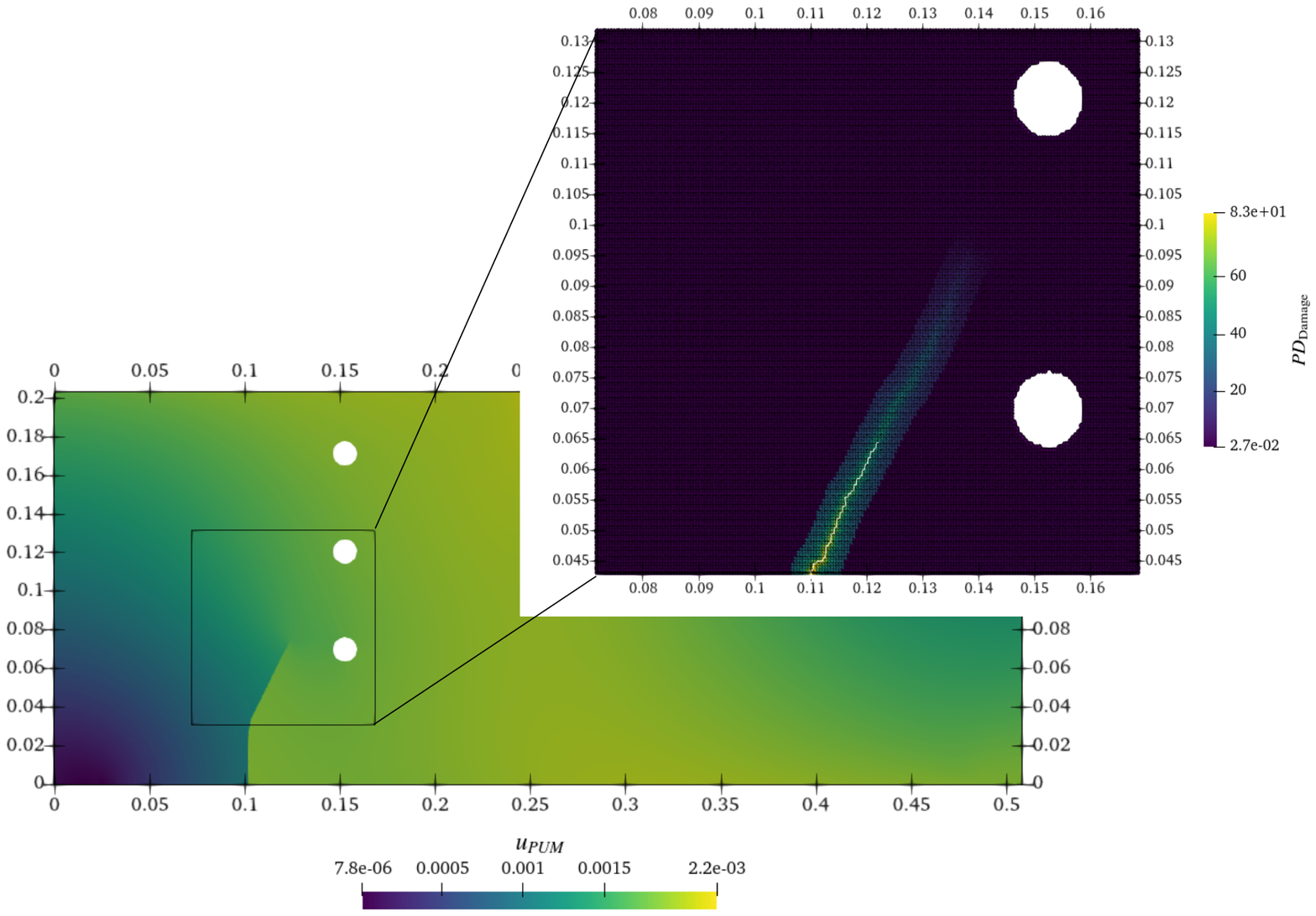}
    \caption{In the background is shown the PUM displacement field for the global problem for Case \Romannum{2} for the second to last load step. The black box is the current box of the sequence of moving PD boxes, see Figure~\ref{fig:3bending:crack:path:case1:box}. In the foreground the final step of the local problem. Here, we show in white the current crack path and in color the PD damage advancing the initial crack path. This path is extracted and fed into the next load step of the global problem as the current crack. Note that the crack path is shown for the reference configuration, not for the deformed configuration.}
    \label{fig:case2:sketch}
\end{figure*}

Before looking into the simulation results for Case \Romannum{2}, we will sketch the combined PUM\textbackslash PD simulator in Figure~\ref{fig:coupling:glcycle} using simulation results. Figure~\ref{fig:case2:sketch} shows the sketch for the second to last load step of the global simulation. In the background, the displacement magnitude of the global PUM simulation is shown for the second to last load step. The black box indicates the PD region. Here, the PUM displacement at the second last load step is applied to the discrete PD nodes within the boundary layer of horizon size $\delta$. In the foreground, the final time step of the PD simulation is shown. The PD nodes are colored with the damage. Where blue means no bonds are softened and yellow means many bonds have softened. The blue line shows the obtained crack from the previous load step. One can see that greenish and yellow nodes after the initial crack. For the next load step of the global problem, the crack path is extracted and transferred to the PUM simulation for the last load step.

Figure~\ref{fig:3bending:crack:path:case2:path} shows in black the experimental crack path for Case \Romannum{2} extracted using \textit{WebPlotDigitizer} from Figure 5.5 in~\cite{ingraffea1990probabilistic}. The dotted black line shows the initial crack. The gray box showcases the PD region used in~\cite{NI2019126} and the green line the simulated crack path using the FEM-PD approach. The green line is close to the experimental crack path. The crack tip moves at the end of the crack and aligns with the experimental crack tip. The red line shows the simulated crack path using our approach. The red line is at the beginning aligned with the experimental crack path. Later the red line diverges from the experimental crack path and the crack tip reaches the void at a different position. Note that we use a similar nodal spacing as in~\cite{NI2019126} for the coarse mesh. However, our PD region is around four times smaller. Figure~\ref{fig:3bending:crack:path:case2:boxes} shows the simulated crack path in red and the moving box in blue. The boxes at the beginning of the simulation are light blue and converge to dark blue over time.

\begin{figure*}[tbp]
    \centering
    \subfloat{
    \begin{tikzpicture}
        \node at (0,0) {Case \Romannum{2}: Number of holes is \num{3}, $a$ = \num{1}, and $b$ = \num{6} };
    \end{tikzpicture}
    }
    \\
    \setcounter{subfigure}{0}
    \subfloat[\label{fig:3bending:crack:path:case2:path}]{
    \begin{tikzpicture}[scale=0.8]
    \begin{axis}[legend pos=south east,xlabel=$x$ distance from centerline of beam (meters), ylabel=$y$ distance from centerline of beam (meters), grid, ymax=0.05,ymin=-4/39.37,xmax=0, xmin=-0.185,xticklabel style={rotate=-45}]
    \addplot[black,mark=none] table [x expr=-(0.254-\thisrowno{0}), y expr=-(0.1016-\thisrowno{1}), col sep=comma] {3point-case2-reference.csv};
    \addplot[asparagus,mark=none] table [x expr=-(0.254-\thisrowno{0}), y expr=-(0.1016-\thisrowno{1}), col sep=comma] {3point-case2-crack-ugo.csv};
    \addplot[awesome,mark=none] table [x expr=-(0.254-\thisrowno{0}), y expr=-(0.1016-\thisrowno{1}), col sep=comma] {3point-case2-crack-64-birner.csv};
    \draw [thick, draw=black,dashed]
    (axis cs: -6/39.37 ,-4/39.37) -- (axis cs: -6/39.37,-3/39.37);
    \draw [thick] (-4/39.37, 0.75/39.37) circle (0.00635);
    \draw [thick] (-4/39.37, -1.25/39.37) circle (0.00635);
    \draw[cadetgrey] (-0.167 , -0.081) rectangle (-0.076 , 0.044);
    \legend{Experiment~\cite{ingraffea1990probabilistic},FEM-PD~\cite{NI2019126},PUM-PD-64};
    \end{axis}
    \end{tikzpicture}
    }
    \subfloat[\label{fig:3bending:crack:path:case2:boxes}]{
    \begin{tikzpicture}[scale=0.8]
    \begin{axis}[legend pos=south east,xlabel=$x$ distance from centerline of beam (meters), ylabel=$y$ distance from centerline of beam (meters), grid, ymax=0.05,ymin=-4/39.37,xmax=0, xmin=-0.185,xticklabel style={rotate=-45}]
    \addplot[awesome,mark=none] table [x expr=-(0.254-\thisrowno{0}), y expr=-(0.1016-\thisrowno{1}), col sep=comma] {3point-case2-crack-64-birner.csv};
    \draw [thick, draw=black,dashed]
    (axis cs: -6/39.37 ,-4/39.37) -- (axis cs: -6/39.37,-3/39.37);
    \draw [thick] (-4/39.37, 0.75/39.37) circle (0.00635);
    \draw [thick] (-4/39.37, -1.25/39.37) circle (0.00635);
    \draw[azure!50] (-0.182 , -0.086) rectangle (-0.086 , -0.015); 
    \draw[azure!65] (-0.182 , -0.079) rectangle (-0.086 , -0.008); 
    \draw[azure!80] (-0.182 , -0.071) rectangle (-0.086 , 0.030); 
    \draw[azure!90] (-0.182 , -0.058) rectangle (-0.086 , 0.030); 
    \legend{PUM-PD-64};
    \end{axis}
    \end{tikzpicture}
    }
    \caption{\protect\subref{fig:3bending:crack:path:case2:path} Experimental obtained crack path for Case \Romannum{2} in~\cite{ingraffea1990probabilistic} in black and the initial crack as a dotted black line. The \textcolor{cadetgrey}{gray} box is the PD domain used in~\cite{NI2019126}. The \textcolor{asparagus}{green} line is the extracted crack path for Case \Romannum{2} using \textit{WebPlotDigitizer} from~\cite{NI2019126}. The \textcolor{red}{line} shows the simulated crack path for a similar nodal spacing as used in~\cite{NI2019126} and the \textcolor{azure}{blue} shows a refined nodal spacing using our approach. \protect\subref{fig:3bending:crack:path:case2:boxes} Simulated crack path with our approach in \textcolor{awesome}{red} and the moving PD region with the crack tip position. The \textcolor{azure!50}{light blue} region at the crack tip is the initial square region. After that, the box color moves from \textcolor{azure!50}{light blue} at the beginning to \textcolor{azure!90}{dark blue} for the region used in the last simulation. For simplicity, we only plot one simulated crack path.}
    \label{fig:3bending:crack:path:case2}
\end{figure*}

\subsection{Case III:}
Figure~\ref{fig:3bending:crack:path:case3:path} shows in black the experimental crack path for Case \Romannum{3} extracted using \textit{WebPlotDigitizer} from Figure 5.7 in~\cite{ingraffea1990probabilistic}. The dotted black line shows the initial crack. The gray box showcases the PD region used in~\cite{NI2019126} and the green line the simulated crack path using the FEM-PD approach. The green line is close to the experimental crack path until the crack kinks. In the middle of the kink, the crack diverges from the experimental crack path and grows to the second void. The red line shows the simulated crack path using our approach. The red line is comparable to the green line until the second void is reached. Here, the green line grows into the void rather straight. The red line shows the curvy shape observed in the experiment. However, both simulation approaches enter the void at a different position as observed in the experiment. We assume that these minor differences can be attributed to the afore mentioned differences in the two coupling approaches and the fact that we employ a dynamic PD simulation.

Figure~\ref{fig:3bending:crack:path:case3:boxes} shows the simulated crack path in red and the moving box in blue. The boxes at the beginning of the simulation are light blue and converge to dark blue over time.

In this example, the initial crack is closer to the voids and is longer as in Case \Romannum{2}, see Table~\ref{tab:geometry:3p:bending}. In part \Romannum{1} of the series, we mention that the choice on how often we exchange the displacement from PUM as boundary conditions to PD and the extracted crack path from PD to PUM is not apriori known. For example for the Case \Romannum{1} with no voids but an initial crack, we could exchange the information every five PUM load step increments. For Case \Romannum{2} with the same initial crack but added voids, this worked quite well and we got comparable results as in the coupling approach in~\cite{NI2019126}.\\

However, in this case, the initial crack is longer and close to the voids. Here, we had to reduce the exchange of information to every load increment from five load increments. For the larger load increments, the crack was attracted by the first void. This behavior has been observed in~\cite{NI2019126} as well. In addition, we had to use a different scheme for running the PD simulation. We had to apply the PUM displacement boundary conditions and let the crack grow for a small increment, update the crack path, and apply the PUM boundary conditions again on the new crack path. This was repeated until the crack was not advancing. A similar approach for quasi-static simulation was presented in~\cite{NI2019126} as scheme B in Figure 5. This concludes if the crack is close to voids or corners of the domain or the boundary, the information should be exchanged more often. However, by using the same scheme to advance the crack, we get comparable results as in~\cite{NI2019126} using a smaller moving box.

\begin{figure*}[tbp]
    \centering
        \subfloat{
    \begin{tikzpicture}
        \node at (0,0) {Case \Romannum{3}: Number of holes is \num{3}, $a$ = \num{1.5}, and $b$ = \num{5} };
    \end{tikzpicture}
    }
    \\
    \setcounter{subfigure}{0}
    \subfloat[\label{fig:3bending:crack:path:case3:path}]{
    \begin{tikzpicture}[scale=0.8]
    \begin{axis}[legend pos=south east,xlabel=$x$ distance from centerline of beam (meters), ylabel=$y$ distance from centerline of beam (meters), grid, ymax=0.05,ymin=-4/39.37,xmax=0, xmin=-0.185,xticklabel style={rotate=-45}]
    \addplot[black, mark=none] table [x expr=-(0.254-\thisrowno{0}), y expr=-(0.1016-\thisrowno{1}), col sep=comma] {3point-case3-reference.csv};
    \addplot[asparagus,mark=none] table [x expr=-(0.254-\thisrowno{0}), y expr=-(0.1016-\thisrowno{1}), col sep=comma] {3point-case3-crack-ugo.csv};
    \addplot[awesome, mark=none] table [x expr=-(0.254-\thisrowno{0}), y expr=-(0.1016-\thisrowno{1}), col sep=comma] {3point-case3-crack-64-birner-new.csv};
    \draw [thick, draw=black, dashed]
    (axis cs: -5/39.37 ,-4/39.37) -- (axis cs: -5/39.37,-2.5/39.37);
    \draw [thick] (-4/39.37, 0.75/39.37) circle (0.00635);
    \draw [thick] (-4/39.37, -1.25/39.37) circle (0.00635);
    \draw[cadetgrey] (-0.133 , -0.070) rectangle (-0.076 , 0.044);
    \legend{Experiment~\cite{ingraffea1990probabilistic},FEM-PD~\cite{NI2019126},PUM-PD-64, new, PUM-PD-128};
    \end{axis}
    \end{tikzpicture}
    
    }
    \subfloat[\label{fig:3bending:crack:path:case3:boxes}]{
    \begin{tikzpicture}[scale=0.8]
    \begin{axis}[legend pos=south east,xlabel=$x$ distance from centerline of beam (meters), ylabel=$y$ distance from centerline of beam (meters), grid, ymax=0.05,ymin=-4/39.37,xmax=0, xmin=-0.185,xticklabel style={rotate=-45}]
    \addplot[awesome,mark=none] table [x expr=-(0.254-\thisrowno{0}), y expr=-(0.1016-\thisrowno{1}), col sep=comma] {3point-case3-crack-64-birner.csv};
    \draw [thick, draw=black, dashed]
    (axis cs: -5/39.37 ,-4/39.37) -- (axis cs: -5/39.37,-2.5/39.37);
    \draw [thick] (-4/39.37, 0.75/39.37) circle (0.00635);
    \draw [thick] (-4/39.37, -1.25/39.37) circle (0.00635);

    \draw[azure!50] (-0.149 , -0.076) rectangle (-0.088 , -0.020); 
    \draw[azure!55] (-0.149 , -0.066) rectangle (-0.088 , -0.012); 
    \draw[azure!60] (-0.142 , -0.066) rectangle (-0.088 , -0.008); 
    \draw[azure!70] (-0.139 , -0.058) rectangle (-0.088 , 0.003); 
    \draw[azure!80] (-0.134 , -0.051) rectangle (-0.089 , 0.005); 
    \draw[azure!85] (-0.134 , -0.043) rectangle (-0.089 , 0.013); 
    \draw[azure!90] (-0.122 , -0.018) rectangle (-0.081 , 0.033); 

    \legend{EPUM-PD-64};
    \end{axis}
    \end{tikzpicture}
    }
    \caption{\protect\subref{fig:3bending:crack:path:case3:path} Experimental obtained crack path for Case \Romannum{3} in~\cite{ingraffea1990probabilistic} in black and the initial crack as a dotted black line. The \textcolor{cadetgrey}{gray} box is the PD domain used in~\cite{NI2019126}. The \textcolor{asparagus}{green} line is the extracted crack path for Case \Romannum{3} using \textit{WebPlotDigitizer} from~\cite{NI2019126}. The \textcolor{red}{line} shows the simulated crack path for a similar nodal spacing as used in~\cite{NI2019126} and the \textcolor{azure}{blue} shows a refined nodal spacing using our approach. \protect\subref{fig:3bending:crack:path:case3:boxes} Simulated crack path with our approach in \textcolor{awesome}{red} and the moving PD region with the crack tip position. The \textcolor{azure!50}{light blue} region at the crack tip is the initial square region. After that, the box color moves from \textcolor{azure!50}{light blue} at the beginning to \textcolor{azure!90}{dark blue} for the region used in the last simulation. For simplicity, we only plot one simulated crack path.} 
    \label{fig:3bending:crack:path:case3}
\end{figure*}

\section{Conclusion}
\label{sec:conclusion}
For fracture mechanics problems, peridynamics is suitable for naturally developing and growing cracks. However, there is no free lunch and the downside is high computational costs. On the other side partition of unity methods can model a known crack path with only a few degrees of freedom. In part I of the series a combined PUM\textbackslash PD simulator was proposed by the authors in~\cite{BIRNER2023103360} to combine the two approaches. Here, we validated the approach against toy problems where we could predict the crack path. The next natural step in this paper was to validate our proposed approach against experimental data. We used the experimental data in~\cite{ingraffea1990probabilistic} to validate our approach. Recall that this experimental data was used to validate peridynamic simulations~\cite{NI2019126,sheikhbahaei2023efficient}, phase-field simulations~\cite{msekh2015abaqus,wu2018robust,mesgarnejad2015validation,gerasimov2020stochastic}, and XFEM/GFEM simulations~\cite{mukhtar2020validation}. We compared our approach with the approach of coupling finite elements with peridynamics in~\cite{NI2019126}. The commonality is a very similar nodal spacing for the PD region the same order of magnitude. In part I, we have shown that the discretization has to be aligned with the initial crack to avoid kinks in the crack path. Thus, we can not use the same nodal spacing. 
Two differences, we want to emphasize are that we use a smaller PD region which moves with the crack, and in~\cite{NI2019126} a bigger constant PD region is used. A smaller box reduces the computational time for the PD region. Notably, our approach using the smaller moving PD region resulted in a very similar crack path for all the experiments as the approach~\cite{NI2019126}. To conclude in part \Romannum{2} of the series, we validated our approach against three experiments and a coupling approach using the finite element method and peridynamics.

However, some steps are still missing for a fully automated combined PUM\textbackslash PD simulator:
\begin{itemize}
    \item Automated identification of the PD region, especially with no initial crack. One possibility is to use some stress-based or strain-based damage model. If the stress or strain exceeds some threshold, the PD region will be placed there. Another choice is the size of the PD region. We observed that small moving regions are sufficient. However, the optimal size of the region to get an accurate crack path while using the box with the smallest computation time is not apriori known. The authors working on a machine learning-based detection of the interface region for coupling local and nonlocal models. In this paper, the authors prescribe the PD domain, but a smaller domain might be sufficient and could reduce the computational cost. The approach for the coupling of nonlocal and local models is suitable for this approach as well.
    \item Automated crack path extraction from the PD damage field. Several algorithms for this are available in the literature~\cite{bussler2017visualization,diehl2017extraction,xfem:pf:nitsche,pf:tip:by:dmg:gradient}. The most promising approach seems to be the \(\theta\)-simplified medial axis algorithm~\cite{sma:extraction:intro,sma:extraction:application} which would construct the medial axis from an iso-contour of the PD damage field. The iso-lines of the damage field can be efficiently extracted by using the marching cubes algorithm~\cite{marching:cubes}.
    \item Once the crack starts to grow
    it is important how often the global and local problem exchange information, \emph{e.g.}\ boundary conditions from the local model to the nonlocal model and the crack path from the nonlocal model to the local problem. In this paper, we observed that for Case \Romannum{3} where the initial crack is much closer to the void, we had to exchange the information more often. In this paper, this was controlled by the authors and needs to be automated.
\end{itemize}
Some additional implementation and research are needed to fully automate the interaction between the PUM and PD code. However, in the current state, our non-fully automated implementation results are in good agreement with the three experimental results presented and comparable with the results in~\cite{NI2019126} where PD was coupled with the finite element method.
Another aspect is three-dimensional simulations. Here, a two-dimensional crack area needs to be extracted as proposed by the authors in~\cite{bussler2017visualization}. However, not many three-dimensional experimental data are available for validation.

\section*{Supplementary materials}
The PeriHPX code~\cite{Diehl2020hpx,Jha2021} utilizing the C++ standard library for concurrency and parallelism (HPX)~\cite{Kaiser2020} is available on GitHub\footnote{\url{https://github.com/perihpx/}}.
PeriHPX has the following dependencies: HPX 1.7.1, Blaze~\cite{iglberger2012high} 3.8, Blaze\_Iterative, YAML-CPP 0.6.3, hwloc 2.2.0, boost 1.73.0, jemalloc 5.2.1, gcc 9.3.0, gmsh~\cite{geuzaine2009gmsh} 4.8, nanoFLANN, and VTK 9.0.1.
The simulation input files are available on Zenodo~\cite{patrick_diehl_2023_10407115} or GitHub\footnote{\url{https://github.com/diehlpk/paperPUMPD-part2}}.
The PUMA\footnote{\url{https://www.scai.fraunhofer.de/en/business-research-areas/meshfree-multiscale-methods/products/puma.html.}} software framework~\cite{schweitzer2017rapid} used for all PUM simulations is developed at Fraunhofer SCAI\@.

\section*{Acknowledgements}

PD thanks the Center of Computation \& Technology at Louisiana State University for supporting this work. This material is partially based upon work supported by the U. S. Army Research Laboratory and the U. S. Army Research Office under Contract/Grant Number W911NF-19-1-0245.

%



\bibliography{references}   

\end{document}